\documentclass[osajnl,twocolumn,showpacs,10pt]{revtex4-1} 
\usepackage{amsmath,amssymb,graphicx}
\usepackage{hyperref} 

\hypersetup{colorlinks=true,
pdftitle={Sub-GHz Faraday filter on the Cs D1 line},
pdfauthor={Mark Zentile},
linkcolor=blue,
citecolor=blue,
urlcolor=blue,
breaklinks=true}

\usepackage[hyphenbreaks]{breakurl}

\newcommand{\mrm}{\mathrm}

\begin{document}

\title{Atomic Faraday filter with equivalent noise bandwidth less than 1 GHz}

\author{Mark A. Zentile}\email{m.a.zentile@durham.ac.uk}
\author{Daniel J. Whiting}
\author{James Keaveney}
\author{Charles S. Adams}
\author{Ifan G. Hughes}
\affiliation{Joint Quantum Center (JQC) Durham-Newcastle, Department of Physics, Durham University, South Road, Durham, DH1 3LE, United Kingdom}

\begin{abstract} We demonstrate an atomic bandpass optical filter with an equivalent noise bandwidth less than 1 GHz using the D$_1$ line in a cesium vapor. We use the ElecSus computer program to find optimal experimental parameters, and find that for important quantities the cesium D$_1$ line clearly outperforms other alkali metals on either D-lines. The filter simultaneously achieves a peak transmission of 77\%, a passband of 310 MHz and an equivalent noise bandwidth of 0.96 GHz, for a magnetic field of 45.3 gauss and a temperature of 68.0$\,^\circ$C. Experimentally, the prediction from the model is verified. The experiment and theoretical predictions show excellent agreement.
\end{abstract}


\maketitle 

\noindent The Faraday effect in atomic media has come to be used for a wide range of applications, including creating macroscopic entanglement~\cite{Julsgaard2001}, GHz bandwidth measurements~\cite{Siddons2009a}, non-destructive imaging~\cite{Gajdacz2013a}, magnetometry~\cite{Behbood2013}, off-resonance laser frequency stabilization~\cite{Marchant2011,Zentile2014}, and creating an optical isolator~\cite{Weller2012d}.

Another application of increasing interest is utilizing the Faraday effect to create ultra-narrow bandwidth optical filters~\cite{Ohman1956}, of the order of a GHz width. These atomic Faraday filters are imaging filters~\cite{Agnelli1975} with a large field of view~\cite{Yeh1982}, and can be engineered to be low loss at the signal frequency~\cite{Fricke-Begemann2002}. This makes them the filter of choice for many applications, for example, they are used in atmospheric lidar~\cite{Chen1996,Fricke-Begemann2002,Huang2009,Harrell2010}, Brillouin lidar~\cite{Popescu2006,Rudolf2012}, Doppler velocimetry~\cite{Cacciani1978,Bloom1993}, free-space communications~\cite{Junxiong1995} and quantum key distribution~\cite{Shan2006}, quantum optics~\cite{Zielinska2014a}, filtering Raman light~\cite{Abel2009}, optical limitation~\cite{Frey2000}, and laser frequency stabilisation~\cite{Sorokin1969,Yabuzaki1977,Wanninger1992,Choi1993,Miao2011}.

Faraday filters have so far been demonstrated experimentally using neon~\cite{Endo1978}, calcium~\cite{Chan1993}, sodium~\cite{Sorokin1969,Agnelli1975,Yabuzaki1977,Chen1993,Hu1998,Yang2011,Kiefer2014}, potassium~\cite{Yin1992,Bloom1993,Billmers1995}, rubidium~\cite{Dick1991,Zielinska2012,Sun2012}, and cesium~\cite{Menders1991,Frey2000,Wang2012}. A Faraday filter on the cesium D$_1$ line (894 nm) could be useful for quantum optics experiments which utilize the the Cs D$_1$ line~\cite{Hockel2010}, and could aid filtering degenerate photon-pairs at 894~nm in a similar way to that shown for 795~nm~\cite{Zielinska2014a}.

In this letter we demonstrate the technique of using computer optimization to find optimal working parameters for a Faraday filter. Using this technique we find that a Faraday filter working at the Cs D$_1$ line has superior performance when compared to similar linear Faraday filters working with different elements and/or transitions. Experimentally, we verify the prediction of the model, and achieve a linear Faraday filter with the best performance to date. 

\begin{figure}[t]
\centerline{\includegraphics[width=\columnwidth]{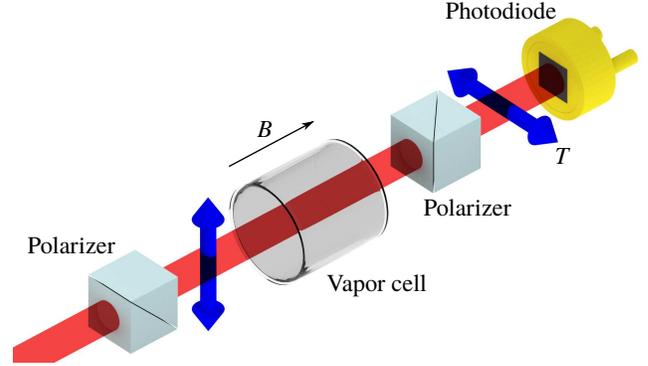}}
\caption{(Color online) Illustration of the experimental arrangement. The vapor cell has an axial magnetic field ($B$) applied, and is placed between two crossed polarizers. The (blue) arrows denote the polarization of the light.}
\label{fig:setup}
\end{figure}

An atomic Faraday filter consists of two crossed polarizers with an atomic vapor cell placed between, as seen in Fig.~\ref{fig:setup}. An axial magnetic field is applied across the vapor cell. This field causes the plane of polarization to rotate when light traverses the cell (the Faraday effect~\cite{Budker2002}) and hence causes transmission through the second polarizer. For an atomic medium the polarization rotation only occurs near atomic resonances (which are intrinsically narrow), resulting in an ultra-narrow filter.

One important measure of filter performance is the equivalent noise bandwidth (ENBW), defined as
\begin{equation}
\mrm{ENBW}=\frac{\int^\infty_0 T(\nu)\mrm{d}\nu}{T(\nu_s)},
\label{eq:ENBW}
\end{equation}
where $T$ the transmission through the filter and $\nu_s$ is signal frequency. In many cases the exact position of the signal frequency can be chosen, and so we take it to be the frequency of maximum transmission ($T(\nu_s) = T_\mrm{max}$). The ENBW is inversely proportional to the signal to noise ratio for narrowband signals in the presence of white noise.

Reducing the ENBW is clearly desirable, though this will come at the cost of reducing the maximum filter transmission~\cite{Kiefer2014}. Many applications require high transmission and as such a good figure of merit (FOM), as proposed in Ref.~\onlinecite{Kiefer2014}, is to use the ratio
\begin{equation}
\mrm{FOM} = \frac{T_\mrm{max}}{\mrm{ENBW}}.
\end{equation}


For a given length of the atomic medium the filter spectrum changes non-trivially as a function of vapor cell temperature and magnetic field. Therefore it is advantageous to have an accurate model of the filter spectrum. We used the ElecSus program~\cite{Zentile2014a} to calculate the filter spectrum. This program calculates the spectrum of the electric susceptibility~\cite{Jackson1999} of an atomic medium for a range of experimental parameters (e.g. magnetic field, cell temperature, cell length, beam polarization etc.), and can output a Faraday-filter spectrum. We interfaced this program with a global optimization algorithm to find the values of temperature and magnetic field that maximize the FOM. Specifically, the random-restart hill climbing meta-algorithm~\cite{Russel2003} was used in conjunction with the downhill simplex method~\cite{Nelder1965} to minimize the inverse of the FOM. This technique used a few thousand evaluations of the filter spectrum from the ElecSus program, taking less than five minutes to complete using a computer with an Intel$^\circledR$ Core\texttrademark ~i3-3220 processor.

Each spectrum is calculated for a range of 60~GHz surrounding the Cesium D$_1$ weighted linecentre, with a 10~MHz grid spacing. The ENBW is then approximated by numerically integrating across the spectrum and then dividing by the maximum transmission value. Note that approximating the ENBW to this frequency range is justified if the optical noise does not extend to other transitions from the Cs ground manifold (such as the D$_2$ line at 852nm), although it is possible to use a Faraday filter in conjunction with an interference filter if this is the case~\cite{Harrell2010}.
\begin{figure}[t]
\centerline{\includegraphics[width=\columnwidth]{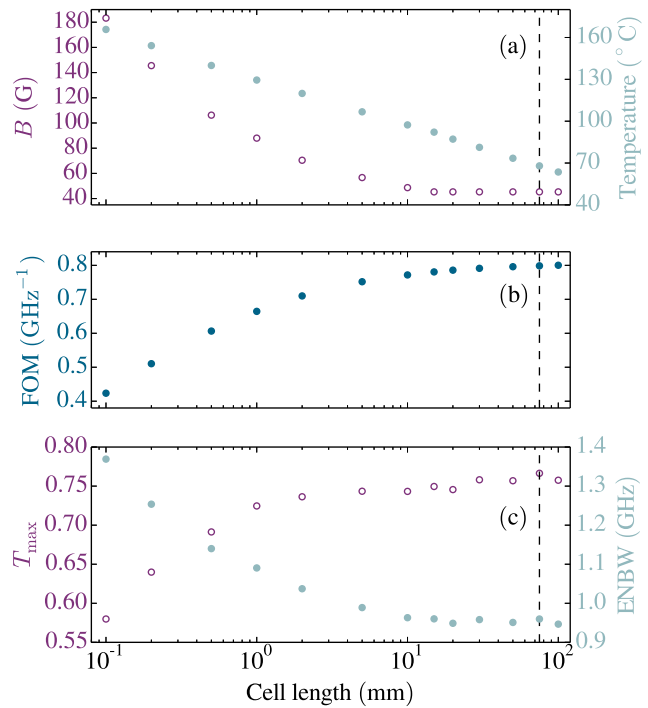}}
\caption{(Color online) Computer optimization of Cs D$_1$ Faraday filters for different cell lengths. Panel (a) shows the values of the parameters required to maximize the figure of merit (FOM); the hollow (purple) circles show the values of magnetic field while the solid (blue) circles give the temperature. Panel (b) shows the optimum FOM values. Panel (c) shows the corresponding ENBW (solid circles) and $T_\mrm{max}$ values (hollow circles). The vertical dashed line marks the length of the experimental cell used to measure the spectrum shown in Fig.~\ref{fig:main}.}
\label{fig:Opt}
\end{figure}
Figure.~\ref{fig:Opt} shows the results of the optimization for vapor cell lengths ranging from 100$\,\mu$m to $10\,$cm. We can see that for cell lengths above 10 mm, the optimal value of the magnetic field is constant at 45.3 G, while the optimal temperature changes such that the change in number density~\cite{Alcock1984} ($\mathcal{N}$) compensates the change in cell length ($L$) in a predictable way; $\mathcal{N}L\approx\mrm{constant}$. However, for cell lengths of 10 mm and less, the optimal magnetic field changes. This is due to self broadening~\cite{Weller2011} becoming important at higher densities, and has a far greater effect than Doppler broadening~\cite{Zentile2015}, as was confirmed by repeating the optimization with the effect of self broadening removed.

To validate the prediction of the model, an experiment using a $75\,$mm long Cs vapor cell was performed. Figure~\ref{fig:setup} shows an illustration of the experimental arrangement. The vapor cell was placed in a solenoid which produced an axial magnetic field and also provided heating of the vapor cell. This was placed between two crossed Glan-Taylor polarizers, forming the Faraday filter. To measure the filter spectrum a continuous wave Ti:sapphire laser was used to produce a beam of light that could be scanned across the Cs D$_1$ line. The weak-probe~\cite{Smith2004,Sherlock2009} beam was sent through the Faraday filter and was detected using an amplified photodiode. The laser scan was calibrated using the method described in Ref.~\onlinecite{Siddons2008}.
\begin{figure}[t]
\centerline{\includegraphics[width=\columnwidth]{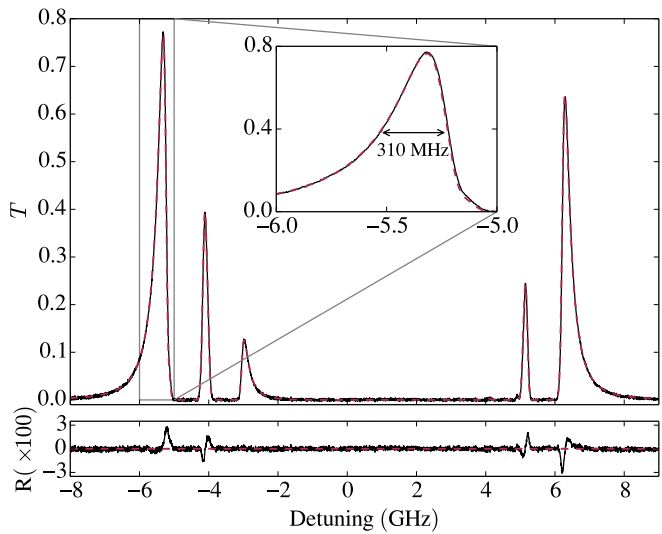}}
\caption{(Color online) Transmission through a Cs D$_1$ Faraday filter with a 75 mm long atomic medium as a function of linear detuning from the weighted linecentre (335.116048807~THz~\cite{Udem1999}). The solid back line shows the experimental data while the dashed (red) line is the theory fit. From the fit, the cell temperature and magnetic field are found to be $(67.8\pm0.3)^\circ$C and $(45.7\pm0.8)\,$G respectively. The ENBW was found to be $(0.96\pm0.01)\,$GHz and the maximum transmission was $(0.76\pm0.01)$. Plotted underneath are the residuals (R) between experiment and theory, with a root mean square (RMS) deviation of 0.4\%, showing excellent agreement.}
\label{fig:main}
\end{figure}

Figure~\ref{fig:main} shows the measured Faraday filter spectrum and a theory fit using ElecSus~\cite{Zentile2014a}. Excellent agreement is found between theory and experiment. The full width at half maximum of the passband was found to be 310 MHz, while the ENBW  and $T_\mrm{max}$ were found to be $(0.96\pm0.01)\,$GHz and $0.76\pm0.01$ respectively; in agreement with the theoretical optimum values. This gives a ratio of $T_\mrm{max}$ to ENBW (FOM) of $(0.794\pm0.015)\,$GHz$^{-1}$, which, to the best of our knowledge, is the highest demonstrated to date for any linear Faraday filter.
\begingroup
\squeezetable
\begin{table}[!t]
  \caption{Prediction of the optimal values of magnetic field ($B$) and temperature (T) by computer optimization for different atoms and transition wavelengths ($\lambda$). Each atomic medium was set to 75~mm long. The Cesium D$_1$ filter, realized in this work (see Fig.~\ref{fig:main}), shows by far the highest ratio of maximum transmission to ENBW (FOM).}
  \begin{center}
    \begin{tabular}{lccccc}
    \hline
    Atom ($\lambda$ [nm])&$B$ [G]& T [$^\circ$C] & ENBW [GHz] & $T_\mrm{max}$ & FOM [GHz$^{-1}$] \\
    \hline
    Na (589) & 128 & 191 & 3.1 & 0.78 & 0.25 \\
    Na (590) & 169 & 198 & 3.3 & 0.80 & 0.24 \\
    K\footnotemark[1] (766) & 76.3 & 107 & 1.8 & 0.80 & 0.44 \\
    K\footnotemark[1] (770) & 92.2  & 113 & 1.8 & 0.82 & 0.45 \\
    $^{85}$Rb (780) & 223 & 54.4 & 2.1 & 0.96  & 0.45 \\
    $^{85}$Rb (795) & 63.8 & 79.8 & 1.1 & 0.75 & 0.66 \\
    $^{87}$Rb (780) & 67.3 & 64.5 & 2.0 & 0.79 & 0.39 \\
    $^{87}$Rb (795) & 61.0 & 82.4 & 1.1 & 0.72 & 0.67 \\
    Rb\footnotemark[1] (780) & 53.4 & 84.1 & 2.3 & 0.91 & 0.40 \\
    Rb\footnotemark[1]\footnotetext[1]{At natural abundance~\cite{Rosman1998}.} (795) & 72.3 & 88.6 & 1.4 & 0.76 & 0.53 \\
    Cs (852) & 55.6 & 47.2 & 1.6  & 0.79 & 0.51 \\
    Cs (894) & 45.3 & 68.0 & 0.96 & 0.77 & 0.80 \\
    \hline
    \end{tabular}
  \end{center}
  \label{tab:75mmOpt}
\end{table}
\endgroup

The computer optimization was also performed for both D-lines of sodium, potassium, rubidium and cesium, for a 75 mm long atomic medium. From Table~\ref{tab:75mmOpt}, we can clearly see that cesium D$_1$ line gives the highest figure of merit. We also find that the performance of rubidium D$_1$ line filters~\cite{Zielinska2012} could be improved further by using an isotopically pure vapor. One thing to note is that, with the exception of $^{85}$Rb at 780 nm, the filter spectra that give the highest FOM value all show `wing' type operation, where the the filter transparencies occur just off-resonance from the atomic transitions. Line-center operation~\cite{Kiefer2014} is found for local FOM maxima. The better FOM for wing operation is explained by the absorption lines cutting the transparencies in the spectra sharply, decreasing the ENBW. 

In conclusion, we have introduced Faraday filtering on the Cesium D$_1$ line. The filter shows excellent performance, surpassing other elements and transitions, and could find use in quantum optics experiments. The method used to find the optimal magnetic field and temperature is fast and simple, utilizing publicly available software~\cite{Zentile2014a} and could be applied for other figures of merit for other applications.

We thank K. J. Weatherill for allowing us use of the Ti:sapphire laser, and C. G. Wade for assistance using the laser.  We acknowledge financial support from EPSRC (grant EP/L023024/1) and Durham University. The data presented in this paper are available from \url{http://dx.doi.org/10.15128/38440c94-7ad2-11e4-b116-123b93f75cba}.

\bibliography{library}

\end{document}